# Predicting Auditory Spatial Attention from EEG using Single- and Multi-task Convolutional Neural Networks

Zhentao Liu[1], Jeffrey Mock[2], Yufei Huang[1], Edward Golob[2]

[1]Department of Electrical and Computer Engineering, [2]Department of Psychology,
the University of Texas at San Antonio, TX 78249

*Abstract*—Recent behavioral and electroencephalograph (EEG) studies have defined ways that auditory spatial attention can be allocated over large regions of space. As with most experimental studies, behavior EEG was averaged over 10s of minutes because identifying abstract feature spatial codes from raw EEG data is extremely challenging. The goal of this study is to design a deep learning model that can learn from raw EEG data and predict auditory spatial information on a trial-by-trial basis. We designed a convolutional neural networks (CNN) model to predict the attended location or other stimulus locations relative to the attended location. A multi-task model was also used to predict the attended and stimulus locations at the same time. Based on the visualization of our models, we investigated features of individual classification tasks and joint feature of the multi-task model. Our model achieved an average 72.4% in relative location prediction and 90.0% in attended location prediction individually. The multi-task model improved the performance of attended location prediction by 3%. Our results suggest a strong correlation between attended location and relative location.

## I. INTRODUCTION

Intelligent behavior requires the ability to both focus spatial attention to help accomplish the current goal while also being responsive to unexpected events at other locations in space. Behavioral and neurophysiological studies suggest that one way spatial attention handles this "dual mandate" is by having a gradient where attentional processing benefits progressively decrease with distance from the attended location [1]. Neurophysiological studies using EEG show that spatial attention gradients likely generated by fronto-parietal brain regions transform absolute spatial locations into a coordinate system centered on the currently attended location [2].

In this paper, we propose a deep convolutional neural network (CNN) model that can extract EEG features that represent various spatial codes operative in an auditory spatial attention task. The CNN model can learn local, lower level features through spatial filters and temporal filters, and then represents higher-level features in the deeper layers [3]. Recent years, CNN models have proven to be successful in many fields such as computer vision, speech recognition, and natural language processing. One useful aspect of CNN models is their effectiveness in end-to-end learning; i.e., learning from raw data without extensive preprocessing and a priori feature selection. This is especially attractive in Brain-Computer Interface (BCI) because it is difficult for humans to select all relevant features from complex EEG signals. We also propose a multi-task learning method to define relations between different tasks and features. Multi-task learning is derived from inductive transfer which can improve network performance by introducing inductive bias [4]. In our case, the inductive bias is contributed by an auxiliary task which causes the model to learn extra features from both tasks to reduce overfitting.

In summary, we proposed individual CNN and multi-task models (MTM) to perform two different predictions in the auditory spatial attention experiment. Participants were told to attend to either a left or right side location while listening to sounds coming from one of five evenly spaced locations within a 180º frontal horizontal plane (45º apart). First, we predicted the sound location about the attended location (i.e. 45º, 90º, 135º, 180º away from the attended location, termed Relative Location Prediction). Second, we predicted where a subject was attending (i.e. left or right side, termed Attended Location Prediction). Each model was interpreted through visualization of learning related features.

## II. EXPERIMENT AND DATA PREPROCESSING

### A. Participants

Forty-four participants were included in this study.

### B. Stimuli

Five virtual white noise burst sounds (0.1–10 kHz, 200 ms duration, 5 ms rise/fall times, ~60 dB nHL) were created to correspond to five locations, each 45º apart in the 180º frontal azimuth plane. The spatialized sounds were created by applying appropriate interaural time and level differences as well as head-related transfer functions for each spatial location. Stimuli were presented with insert earphones rather than free-field speakers in order to limit the influence of visual indicators of sound sources and avoid changes in the relationship between sound source location and the ears due to head movements.

### C. Experimental Paradigm

A schematic of the paradigm with a sample stimulus sequence is shown in Figure 1. All participant were given a response pad and told to listen to sequences of white noise coming from one of five possible locations in the frontal azimuth plane (left to right: -90°, -45°, 0°, +45°, +90°). Within each block, a participant was given a target location (left or right, in separate blocks) and told to respond as quickly as possible while ensuring accurate responses to a designated target location by pressing a button with the thumb of their dominant hand. Each location had a .20 probability (target p = 0.20, non-target p = 0.80, .20 probability/non-target) and was randomly presented within each block (stimulus onset asynchrony = 2.4 s). There were 150 stimuli per block. Each target location had two blocks for a total of four blocks. Behavioral measures to targets included median reaction time, hit rate (percentage of responses to target) and false alarm rate (percentage of responses to non-targets).

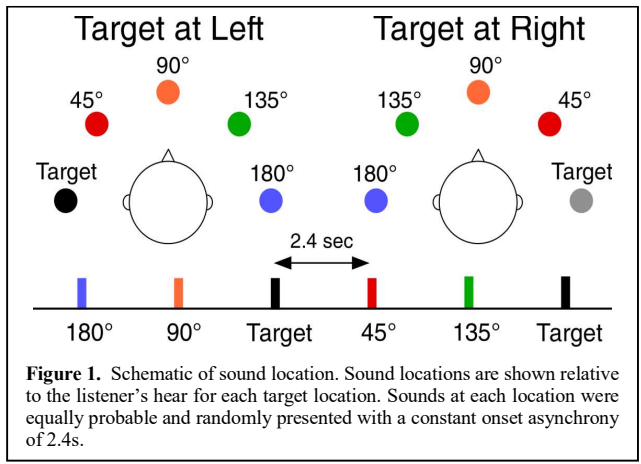

**Figure 1.** Schematic of sound location. Sound locations are shown relative to the listener's hear for each target location. Sounds at each location were equally probable and randomly presented with a constant onset asynchrony of 2.4s.

*D. EEG Recording*

All EEG data was recorded in a sound-attenuated booth by a 64 channels electrode cap, and continuously digitized at 500 Hz with 64-channel EEG system (Compumedics Neuroscan) and stored for off-line analysis.

*E. Data preprocessing*

In this study, we minimally preprocessed the EEG datasets to maximize CNN end-to-end learning ability. We rejected bad channels by standard deviation threshold [-2,2] and used spherical interpolation to make new channels. Data were filtered (1-45 Hz) and then epoched (-1.2 to 1.2 s, relative to sound onset). After preprocessing we obtained 25,975 samples, and each sample had 64 channels and 350 time points. All preprocessing work was done on EEGLAB which is an open source software toolbox.

## III. METHOD

*A. Convolutional neural networks*

In general, convolutional neural networks for EEG signals learn local features through convolution kernels and transforms high dimensional features into lower dimensions that contain global information about the original data [5].

In this study, we used a blocked design for our CNN model. In Block 1 we apply two convolutional kernels in sequence. First, we used kernels of size (channels, 1) to learn spatial correlation; with 64 weights for all 64 channels. We then used kernels of size (1, C) to learn temporal correlations. Hyperparameter C was determined during hyperparameter optimization. The two convolutions were kept linear because there is no significant performance improvement for nonlinear activation [6]. After convolution operations, we performed Batch Normalization and Exponential Linear Unit (ELU) activation [7]. Then a maxpooling layer was added to reduce the dimension of features. Regularization used both L1-L2 norm with dropout to limit overfitting.

Block 2 has a similar structure to block1 except spatial filters are not included. Block 3 and block 4 have the Block 2 structure, and will be added to fit different tasks.

The classification block collects low dimension features directly after the last convolution block. Activation functions are set up as Sigmoid for binary classification, with SoftMax for multi-label classification.

*B. Relative location prediction*

In this task, we refer to the relative location as the distance between the attended location and a given sound's location. When subjects attend to the left, speakers )left to right: -90°, -45°, 0°, 45°, 90°) will be labeled as 0, 1, 2, 3, 4, respectively. When the attended location is on the right, labels will be (left to right) 4, 3, 2, 1, 0. Thus, the model will classify EEG data into these 5 classes without knowing the absolute location of the attended location (code=0 for both left and right attention).

We use four convolution blocks in this task; each block has the same dropout rate of 0.6 and regularization rate of 0.001. We choose the first three blocks with a temporal filter size of (1,10), and the last block as (1,6) due to epoch length. In compiling we use optimizer Adam with an adaptive learning rate of 0.01 and decay=0.001 in every epoch during 400 training iterations. All parameters and hyperparameters are chosen by Hyperas grid searching algorithm. The architecture is shown in Table 1.

After the model is trained, we extract all spatial filters, each of them is a (64,1) vector and the elements are the weights the model learned for each channel. We mapped those weights back on to the scalp sites to study define signal topography. We assume that the most significant filter provides the most discriminative features, so the filters are ranked based on feature classification performance to find the best spatial features.

We also performed slope analysis on this task. Each of the feature maps for spatial filters is extracted as an ERP map, and

| Table 1. Relative Location Model ||||
|---|---|---|---|
| Layer | Output Shape | Layer | Output Shape |
| Input Layer | 64, 350, 1 | Batch Norm. | 1, 25, 100 |
| Conv2d | 1, 350, 10 | ELU | 1, 25, 100 |
| Conv2d | 1, 341, 25 | Max Pooling | 1, 8, 100 |
| Batch Norm. | 1, 341, 25 | Dropout | 1, 8, 100 |
| ELU | 1, 341, 25 | Conv2d | 1, 3, 200 |
| Max Pooling | 1, 113, 25 | Batch Norm. | 1, 3, 200 |
| Dropout | 1, 113, 25 | ELU | 1, 3, 200 |
| Conv2d | 1, 104, 50 | Max Pooling | 1, 1, 200 |
| Batch Norm. | 1, 104, 50 | Dropout | 1, 1, 200 |
| ELU | 1, 104, 50 | Flatten | 200 |
| Max Pooling | 1, 34, 50 | Dense | 5 |
| Dropout | 1, 34, 50 | Softmax | 5 |
| Conv2d | 1, 25, 100 | | |
| Total params: 187,046 ||||
| Trainable params: 187,038 ||||

then the linear slope of the amplitude across the 4 non-target locations was calculated at each time point.

*C. Attended location prediction*

In this task, we want to predict the subject's attended location based on EEG signals. From a previous study using averaged EEG signals the gradient away from left and right attended locations was very similar. This posed a challenge for the model to tell the difference between attending to the left and right side. Our solution was to feed the sound location along with the EEG signal to the model using label embedding so that model can use one more factor to make a prediction. Absolute locations are labeled as 1, 2, 3, 4, 5 from left to right as another input for the model.

We used a similar architecture with the relative location prediction task, but had three blocks instead of four. Each block has a dropout rate of 0.5 and regularization rate of 0.001. All three blocks of temporal filter's size is (1,10). We added one embedded layer to model to expand sound location from a single number to a vector that is also learned by the model. We merge this vector with low dimension features from the top convolution layer, then send the merged feature to the classifier. For compiling we used the optimizer Adam with an adaptive learning rate of 0.01 and decay=0.001 for every epoch (600 training iterations). All parameters and hyperparameters were chosen by the Hyperas grid searching algorithm. The architecture is shown in Table 2.

After the model was trained, we extracted all spatial filters from the first block, and the weights were mapped onto the scalp to visualize their topography. As in this task, we do not have 5 classes to label, we use Elastic Net to do a "one stone two birds" classification task. In Elastic Net regression 350 time points were treated as features, and the algorithm will assign a coefficient for each of the points [8]. From the coefficients, we can construct a heatmap to show significant time points from all 350 features. At meantime, regression performance could give us the spatial filter ranking which is similar to what we did in the previous task.

*D. Multi-task model (MTM)*

An MTM was used to combine the attended location task and relative location task in one model. The multi-task model learned features from both the attended and relative locations, which were expected to improve model performance on each task. However, the attended location prediction needed to take sound location for input. Consequently, the shared feature learned by model contains absolute location information while relative location prediction represents the distance between sound and attended locations. Use of absolute and relative coding can pose a challenge. For example, in the relative location task, the model does not distinguish location 0°left attending) and location 0° (right attending) because the distances from those two locations to target are both zero, hence they should have the same label. However, in the attended location task, these two locations do have two different labels (left vs. right). Therefore, we mainly focus on improving attended location task performance rather than both the tasks.

Our multi-task model contains three blocks as a shared feature extractor while keeping two task-specific output layers. We added one dense layer before attended location task classifier for conveniently merging the embedded labels. The architecture is shown in Table 3. The loss function $L_{total}$

**Table 2.** Attended Location Model

| Layer | Output Shape | Layer | Output Shape |
|---|---|---|---|
| Input Layer_1 | 64, 350, 1 | Conv2d | 1, 25, 100 |
| Conv2d | 1, 350, 10 | Batch Norm. | 1, 25, 100 |
| Conv2d | 1, 341, 25 | ELU | 1, 25, 100 |
| Batch Norm. | 1, 341, 25 | Max Pooling | 1, 8, 100 |
| ELU | 1, 341, 25 | Dropout | 1, 8, 100 |
| Max Pooling | 1, 113, 25 | Flatten | 800 |
| Dropout | 1, 113, 25 | Input Layer_2 | 1 |
| Conv2d | 1, 104, 50 | Embedding | 1, 800 |
| Batch Norm. | 1, 104, 50 | Flatten | 800 |
| ELU | 1, 104, 50 | Multiply | 800 |
| Max Pooling | 1, 34, 50 | Dense | 2 |
| Dropout | 1, 34, 50 | Sigmoid | 2 |
| Total params: 72,239 Trainable params: 72,233 | | | |

**Table 3.** Multi-task model

| Layer | Output Shape | Layer | Output Shape |
|---|---|---|---|
| Input Layer_1 | 64, 350, 1 | ELU | 1, 25, 120 |
| Conv2d | 1, 350, 15 | Max Pooling | 1, 8, 120 |
| Conv2d | 1, 341, 30 | Dropout | 1, 8, 120 |
| Batch Norm. | 1, 341, 30 | Flatten | 960 |
| ELU | 1, 341, 30 | Dense | 5 |
| Max Pooling | 1, 113, 30 | Softmax | 5 |
| Dropout | 1, 113, 30 | Input Layer_2 | 1 |
| Conv2d | 1, 104, 60 | Embedding | 1, 200 |
| Batch Norm. | 1, 104, 60 | Flatten | 200 |
| ELU | 1, 104, 60 | Dense | 200 |
| Max_pooling2d | 1, 34, 60 | Multiply | 200 |
| Dropout | 1, 34, 60 | Dense | 2 |
| Conv2d | 1, 25, 120 | Sigmoid | 2 |
| Batch Norm. | 1, 25, 120 | | |
| Total params: 294,304 Trainable params: 294,298 | | | |

of MTM is a linear combination of the above two tasks $L_i$ as Eq (1) [9].

$$L_{total} = \sum \alpha_i L_i \quad (1)$$

In the model selection process, we put coefficients $\alpha_1$ and $\alpha_2$ in Hyperas searching space so that we can search for a relatively good linear loss combination to maximize the model performance.

## IV. RESULT AND DISCUSSION

### A. Relative location task

Model performance for relative location is shown in Figure 2. As our data are balanced in 5 labels, we use one vs all ROC_AUC to evaluate performance. Result show class 0 (attended target) has the best prediction, and likely reflects attention bias specific to targets. On the other hand, class 2 (90° direction) is quite difficult due to its neutral position.

We extracted features associated with the 10 spatial filters and performed classification (Table 4). Spatial filter 9 had the best performance and was the most discriminative filter that produced significant features. The topography spatial filter 9 is shown in Figure 3. From the topography, the central frontal cluster exhibit high activity in relative location classification. Event-related potentials for each class were from generated from the spatial filters, and all five classes show a positive potential (P300) at a latency of ~500 ms. Event-related potentials from Spatial filter 9 have three clear sensory peaks (~100-200 ms), as well as positivity at ~440-550 ms after sound onset that reflected attention gradients. We further performed attention gradient analysis by fitting the linear slope of potentials across the 4 non-target locations (350 time points and slope values) in Figure 4. The slope over time curve corresponds to significant EEG gradients discovered in previous work [2].

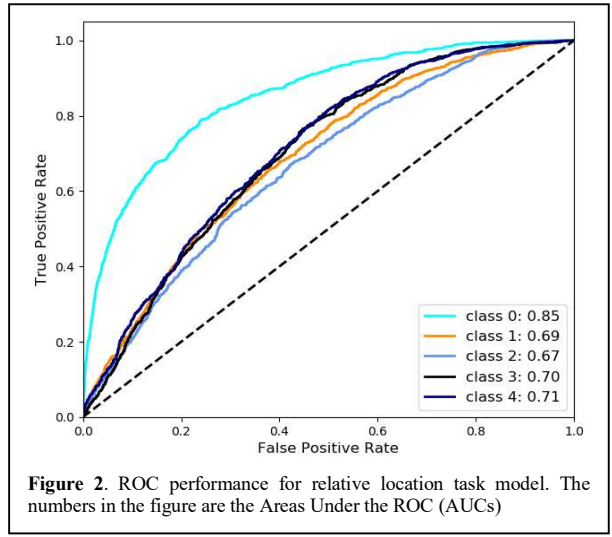

**Figure 2**. ROC performance for relative location task model. The numbers in the figure are the Areas Under the ROC (AUCs)

| Table 4. Spatial filter performance (relative location) | | | | |
|---|---|---|---|---|
| Spatial1 | Spatial2 | Spatial3 | Spatial4 | Spatial5 |
| 60.40% | 52.07% | 60.10% | 61.81% | 57.59% |
| Spatial6 | Spatial7 | Spatial8 | Spatial9 | Spatial10 |
| 57.46% | 53.45% | 49.29% | 67.50% | 49.97% |

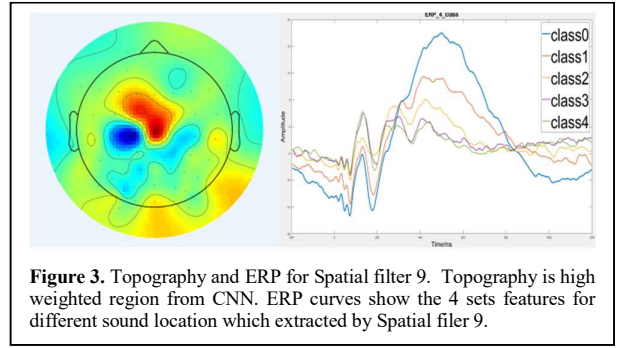

**Figure 3.** Topography and ERP for Spatial filter 9. Topography is high weighted region from CNN. ERP curves show the 4 sets features for different sound location which extracted by Spatial filer 9.

### B. Attended location prediction

Model performance is shown in Figure 5. With the embedded label's help, attended location prediction achieved 90% ROC_AUC performance. Similar to the relative location task, we extracted features from 10 spatial filters and performed classification. Unlike the previous task, the attended location only had 2 labels (left, right). Thus we did not perform gradient analysis. Instead, we used Elastic Net Regression for feature selection. The regression performance ranking results are in Table 5. Spatial filters 1, 3, 4, and 9 were selected due to their high ranking. In the Elastic Net, the regression function follows Eq (2) [8].

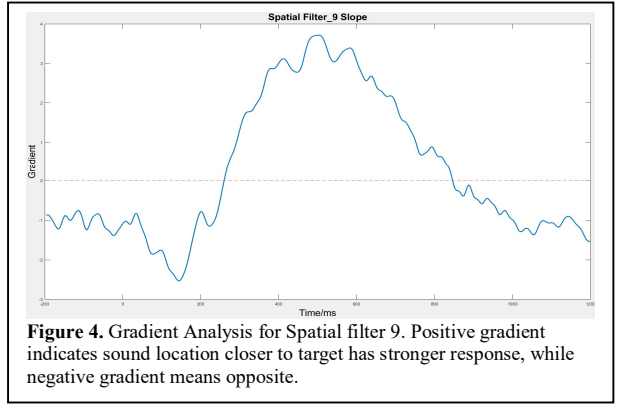

**Figure 4.** Gradient Analysis for Spatial filter 9. Positive gradient indicates sound location closer to target has stronger response, while negative gradient means opposite.

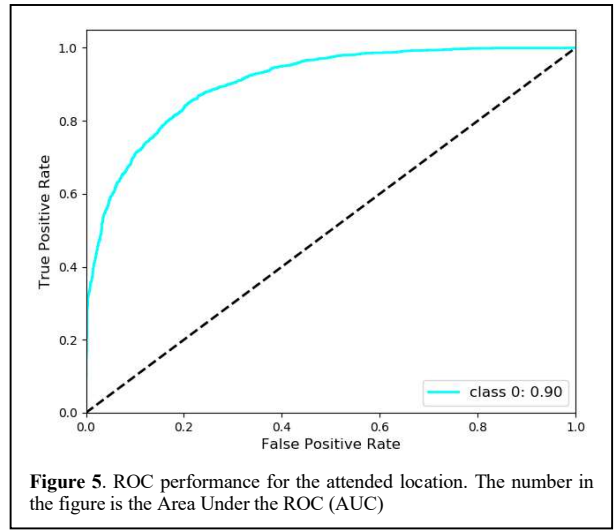

**Figure 5**. ROC performance for the attended location. The number in the figure is the Area Under the ROC (AUC)

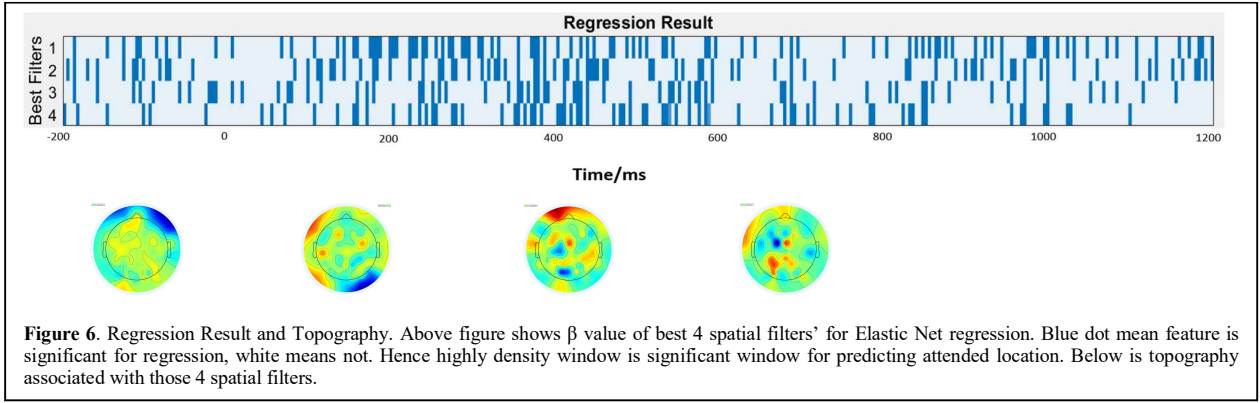

**Figure 6.** Regression Result and Topography. Above figure shows β value of best 4 spatial filters' for Elastic Net regression. Blue dot mean feature is significant for regression, white means not. Hence highly density window is significant window for predicting attended location. Below is topography associated with those 4 spatial filters.

**Table 5.** Spatial filter performance (attended location)

| Spatial1 | Spatial2 | Spatial3 | Spatial4 | Spatial5 |
|---|---|---|---|---|
| 20.31% | 21.66% | 22.23% | 25.60% | 22.52% |
| Spatial6 | Spatial7 | Spatial8 | Spatial9 | Spatial10 |
| 22.14% | 25.12% | 23.29% | 26.66% | 22.91% |

$$\beta = \mathrm{argmin}\ (\|y - X\beta\|^2 + \lambda_2 \|\beta\|^2 + \lambda_1 \|\beta\|_1) \quad (2)$$

Hence each feature will be assigned one β and we can use β value to evaluate the feature's significance level. Heatmap of β which associates with 4 spatial filters is shown in Figure 6. It demonstrates that the most significant time window is between ~100-600 ms. Not every filter had a clear topographic pattern, which is likely due to noise from other source affecting the minimum preprocessing method. Even with noise, the model still had 90% AUC performance.

### C. Multi-task model

In our Multi-task model the attended location and relative location are predicted at the same time. One top consideration for multi-task models is the joint loss function construction. Due to different ranges of individual task loss functions, simply adding them together will cause performance collapse for a small loss range task [10]. In our case, the embedded label in attended location prediction may be underdetermined. For example, an embedded label shows a significant difference between sound from left target and sound from the right target, but in relative location, their label should be both class 0. The goal is simply to use relative location as a constraint to improve other task performance. Therefore, we do not emphasize the relative location prediction. We used a linear combination of individual loss function for the multi-task model joint loss function. Several pairs of weights that favor the attended location prediction are put into Hyperas searching grid for optimization, (0.6, 0.4) was selected eventually. Two tasks ROC_AUC performances are shown in Figure 7.

The attended location prediction had a significant 3% performance improvement, while relative location prediction dropped a little bit. However, in relative location prediction, class 0 still has the best performance among all locations while class 2 is the most confusing direction for the model. The trend of performance maintained well as shown in the individual model. That implied, relative location task was not collapsed. The feature selection mechanism is working properly.

To further explore what features are captured in MTM, we extracted the samples correctly classified in MTM but misclassified in the individual attended location task model. In total there are 457 (183 at -90° location, 274 at +90° location) of them. Misclassified samples were passed to spatial filters, then took average in respect of numbers of filters. Therefore we obtain weighted ERP feature maps. By comparing them with the samples are classified correctly in individual attended location task, we would be able to identify the extra feature captured by MTM. The ERP heatmap is shown in Figure 8.

Note that differential samples have a lower amplitude in general, and the time window was much wider. From the relative location task, we learned that in the 440-550 ms time window sounds closer to attended location had higher responses (depicted in Figure 9). Hence the nature of predicting attended location is to tell the difference between two curves.

Therefore, it makes sense when low amplitude samples have worse performance in the individual model. When the sample has lower amplitude, the gradient of different locations response also becomes lower. As a result, the curves shown in Fig 11 would be flatter in general which makes them more difficult to be classified correctly. Other study implied if the coming target sound showed up right after previous target sound, the amplitude of P300 response will be much lower than usual, which is referred as sequence effect. We believe the sequence effect played a significant role in this task. The number of misclassified samples also approximately matched the data suffered from sequence effect, around 10% out of 5195. These samples are classified correctly in MTM, because the feature selected by relative location task is supposed to have a stronger ability to distinguish amplitudes for steeper slope gradients, which would enhance the difference of attended location curve.

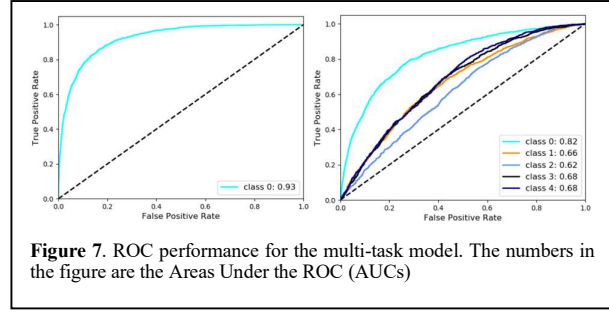

**Figure 7.** ROC performance for the multi-task model. The numbers in the figure are the Areas Under the ROC (AUCs)

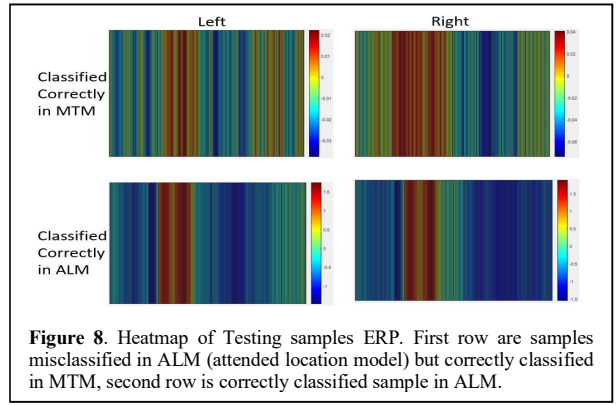

**Figure 8.** Heatmap of Testing samples ERP. First row are samples misclassified in ALM (attended location model) but correctly classified in MTM, second row is correctly classified sample in ALM.

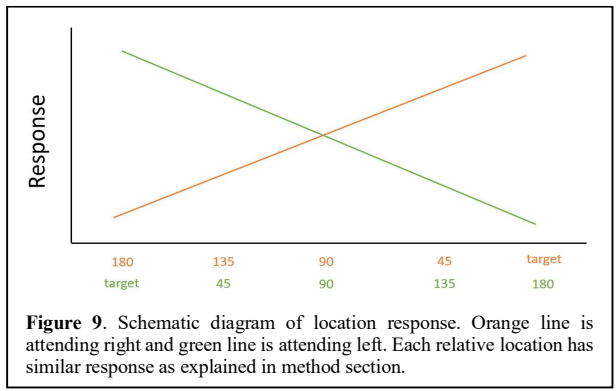

**Figure 9**. Schematic diagram of location response. Orange line is attending right and green line is attending left. Each relative location has similar response as explained in method section.

## V. CONCLUSION AND FUTURE WORK

This study applies CNN models to predicting stimulus and attended locations from EEG recorded during in an auditory spatial attention experiment. Our classification results based on single trial event-related potential showed good performance on both individual task model and multi-task mode. We interpreted our results with CNN visualization method, which not only confirmed auditory spatial attention features form previous work but also successfully found new joint features.